\documentclass[a4paper,11pt]{article}
\pdfoutput=1
\usepackage[utf8]{inputenc}
\usepackage{jheppub}
\usepackage{amssymb}
\usepackage{mathtools}
\usepackage{physics}
\usepackage{graphicx}
\usepackage{enumitem}
\usepackage{hyperref}
\usepackage{slashed}
\usepackage{orcidlink}
\usepackage{bbm}

\newcommand{\be}{\begin{equation}}
\newcommand{\ee}{\end{equation}}
\newcommand{\bea}{\begin{eqnarray}}
\newcommand{\eea}{\end{eqnarray}}

\usepackage{color}
\usepackage[normalem]{ulem}

\def \Nf{N_\textmd{f}}

\graphicspath{{.}{../Figs/}}

\begin{document}
\title{Crossing the desert: Towards predictions for SMEFT coefficients from quantum gravity
}
\author[a]{Lydia Brenner\orcidlink{0000-0001-5350-7081},}
\author[b]{Abhishek Chikkaballi\orcidlink{0000-0003-2240-528X},}
\author[c]{Astrid Eichhorn\orcidlink{0000-0003-4458-1495},}
\author[c]{Shouryya Ray\orcidlink{0000-0003-4754-0955}}

\affiliation[a]{Nikhef, Science Park 105, 1098 XG Amsterdam, The Netherlands}
\affiliation[b]{National Centre for Nuclear Research, Pasteura 7, 02-093 Warsaw, Poland}
\affiliation[c]{CP3-Origins,  University  of  Southern  Denmark,  Campusvej  55,  DK-5230  Odense  M,  Denmark}

\emailAdd{lbrenner@nikhef.nl}
\emailAdd{abhishek.chikkaballiramalingegowda@ncbj.gov.pl}
\emailAdd{eichhorn@cp3.sdu.dk}
\emailAdd{sray@cp3.sdu.dk}

\abstract{The SMEFT provides a general framework to search for new physics beyond the current reach of direct detection. One such form of new physics is quantum gravity. Based on dimensional analysis, one would expect the prediction that the quantum-gravity contribution to the SMEFT coefficients is unmeasurably tiny at LHC scales. In this paper, we test this expectation in a specific framework for quantum gravity, namely the asymptotic safety framework. In this framework, Wilson coefficients can be calculated in relatively straightforward manner, making a connection between quantum gravity and LHC tests of the SMEFT achievable. We work in a toy model of the Standard Model fermion sector to investigate four-fermion couplings. We find three scenarios in this toy model, based on three distinct fixed points of the Renormalization Group flow. In the first scenario, the expectation from dimensional analysis is borne out and Wilson coefficients are Planck-scale suppressed. In the second and third scenarios, the Wilson coefficients are significantly larger than expected by dimensional analysis, due to interacting fixed points which generate an effective new-physics scale that lies between the LHC scale and the Planck scale. We comment on the implications of these results for the testability of asymptotically safe gravity within the SMEFT framework at the LHC.
}
	
\maketitle

\section{Introduction and Motivation}
\label{sec:intro}
Given the absence of direct evidence for new physics at the LHC, the search for indirect evidence is gaining importance. A powerful pathway to search for the imprints of new physics is through the framework of the Standard Model Effective Field Theory (SMEFT) \cite{Buchmuller:1985jz,Grzadkowski:2010es}, see \cite{Ellis:2021kzk,Falkowski:2023hsg} for reviews and \cite{Hartland:2019bjb, Brivio:2019ius, Biekotter:2018ohn, Ellis:2018gqa, daSilvaAlmeida:2018iqo, Aebischer:2018iyb, Ellis:2020unq,Bruggisser:2021duo, Bruggisser:2022rhb,ATLAS:2024lyh,Falkowski:2017pss,terHoeve:2023pvs} for an overview of experimental constraints from the LHC. The SMEFT contains all quasi-local interactions of the Standard Model (SM) fields that are compatible with the SM symmetries, organized by power counting. The coefficients of these interactions change in the presence of new physics. New physics at the mass scale $\Lambda_{\text{NP}}$ is expected to generate dimension-$d$ operators suppressed by $1/\Lambda_{\text{NP}}^{d-4}$. 
Thus, dimension-five and dimension-six operators are expected to be the strongest probes of new physics, if we assume that the dimensionless coefficients are all numbers of the same order. Given a particular new-physics model, the higher-order Wilson coefficients in the SMEFT can be calculated\footnote{This assumes that standard power-counting holds and higher-order couplings do not introduce new free parameters, as they in principle can do, e.g., in strongly-coupled settings.}. A confrontation with the experimental bounds on the SMEFT coefficients then allows one to rule out or constrain the new-physics model, even if its degrees of freedom lie at higher masses than the center-of-mass-energy at the LHC and on-shell production is therefore not possible.

There is new physics that we already know about: gravity, which is not really new physics, but rather very well known or ``old" physics, but which is challenging to include in the SM. Different candidate theories of quantum gravity exist that provide different descriptions of the fundamental nature of spacetime and its quantum degrees of freedom and interactions with matter.
Distinguishing them through experimental results is of paramount importance. In this paper, we ask whether we can use the SMEFT as a framework to constrain this form of new physics.  In this framework, an effective Lagrangian $\mathcal{L}_{\text{EFT}}$ is defined by the SM Lagrangian $\mathcal{L}_{\text{SM}}$, supplemented by additional higher-order operators, such as dimension-6 operators $\mathcal{O}^{(6)}$ and dimension-8 operators $\mathcal{O}^{(8)}$,
\begin{equation}
     \mathcal{L}_{\text{EFT}} = \mathcal{L}_{\text{SM}} + \sum_i{\frac{c_i}{\Lambda_{\text{NP}}^2}\mathcal{O}_i^{(6)}} + \sum_j{\frac{c_j}{\Lambda_{\text{NP}}^4}\mathcal{O}_i^{(8)}} + \ldots
     \label{eq:smeftlagrangian}
\end{equation}
Here, the $c_i$ specify the strengths of the new interactions and are known as the Wilson coefficients, and $\Lambda_{\text{NP}}$ is the scale of new physics. 
At a first glance, one would expect that no information on quantum gravity can be gleaned from the SMEFT framework, because the SMEFT coefficients are expected to be $\frac{c}{\Lambda_{\text{NP}}^n}$, where $c = \mathcal{O}(1)$ and $n$ grows with the mass-dimension of the operator. For $\Lambda_{\text{NP}} = M_{\rm Planck} \approx 10^{19}\, \text{GeV}$, we expect the SMEFT coefficients to be much too small to be detectable at the LHC, even for dimension-5 (6) operators with $n=1$ ($n=2$). This, however, assumes that $c = \mathcal{O}(1)$ holds. This is an assumption that we investigate in this paper. We work in the framework of asymptotically safe quantum gravity, see \cite{
Eichhorn:2018yfc, Eichhorn:2020mte,Bonanno:2020bil,Reichert:2020mja,Pawlowski:2020qer,Eichhorn:2022jqj,
Eichhorn:2022gku,Saueressig:2023irs,Eichhorn:2023xee,Pawlowski:2023gym} for recent reviews and lecture notes, which very naturally lends itself to a connection with the SMEFT for several reasons. First, it is a quantum field theory, and therefore has the same field content as the SMEFT, plus the metric field. Second, it is typically investigated in the framework of the effective action, which automatically contains the same higher-order interactions as the SMEFT. In fact, one may relate the couplings in a suitable expansion of the effective action quite directly to the Wilson coefficients in the SMEFT.\footnote{This relation of couplings to Wilson coefficients was recently used to for the first time connect the predictions from asymptotically safe gravity to positivity bounds for four-photon-interactions \cite{Knorr:2024yiu,Eichhorn:2024wba}.} 

In the present paper, we investigate four-fermions couplings. For the SMEFT, the LHC has already constrained the Wilson coefficients of a subset of four-fermion couplings \cite{ATLAS:2021kog,Collaboration:2741341, Bellan:2021dcy}. In order to prepare a direct confrontation  of quantum gravity theories with experimental data in future work, we work with a toy-model here which does not account for all SM degrees of freedom. Our toy model does, however, have sufficient structure to exhibit three different scenarios regarding quantum-gravity predictions for four-fermion couplings.\\
We work under the assumption of a ``desert" between the Planck scale and the LHC scale, i.e., we assume that other new physics besides gravity may exist, but does not significantly impact four-fermion couplings. The situation becomes more subtle if this assumption does not hold and another new-physics scale $\Lambda_{\rm NP'}< \Lambda_{\text{NP}}$ exists, because then any predictions from quantum gravity at $\Lambda_{\text{NP}}$ are likely ``washed out" by the physics at $\Lambda_{\rm NP'}$. Such a ``desert" assumption is compatible with new physics in the neutrino sector and with many dark-matter models, with one example being \cite{Asaka:2005pn}.

This paper is structured as follows: To introduce the respective other research communities to the topic, we review the experimental status of four-fermion operators in Sec.~\ref{sec:reviewSMEFT} and asymptotically safe gravity-matter models in Sec.~\ref{sec:review}. In Sec.~\ref{sec:fourfermionresults} we introduce the toy model of SMEFT-four-fermion operators, present the beta functions and fixed-point structure. In Sec.~\ref{sec:scenarios} we describe three scenarios for the values of four-fermion couplings at the LHC which can all be realized in our toy model. Finally, in Sec.~\ref{sec:conclusions} we discuss potential implications of these scenarios for how LHC measurements could provide information on quantum gravity.

\section{Review: Experimental constraints on four-fermion operators in the SMEFT}
\label{sec:reviewSMEFT}
In general, analyses at the LHC aim to constrain the dimension-6 Wilson coefficients that correspond to operators that either
directly or indirectly impact particle couplings. Contributions of operators of mass-dimension 8, which are suppressed by $\frac{1}{\Lambda_{\text{NP}}^2}$ relative to the leading effects from dimension 6 operators and whose impact on couplings in the kinematic regions of interest are not fully calculated, are not considered. Furthermore, a value of $\Lambda_{\text{NP}} = 1$ TeV is generally assumed. Coefficients for alternative values of $\Lambda_{\text{NP}} = X$ can be obtained through a scaling of the results presented in the measurements by a factor $(X/1 \textsc{TeV})^2$. All Wilson coefficients are assumed to be real. The 4-fermion dimension-6 SMEFT operators in the Warsaw basis \cite{Grzadkowski:2010es} are depicted in Table \ref{tab:dim6wilsonfermi}.

\begin{table}[]
    \centering
    \begin{tabular}{c|c || c|c || c|c}
      $c_i$   & Operator &  $c_i$   & Operator & $c_i$  & Operator\\
      &$(\bar{L}L)(\bar{L}L)$&&$(\bar{R}R)(\bar{R}R)$&&$(\bar{L}L)(\bar{R}R)$\\
      \hline
       $c_{ll}$  & $(\bar{l}_p \gamma_{\mu}l_r)(\bar{l}_s\gamma^{\mu} l_t)$ &  $c_{ee}$  & $(\bar{e}_p \gamma_{\mu}e_r)(\bar{e}_s\gamma^{\mu} e_t)$ &  $c_{le}$  & $(\bar{l}_p \gamma_{\mu}l_r)(\bar{e}_s\gamma^{\mu} e_t)$\\
        $c_{qq}^{(1)}$  & $(\bar{q}_p \gamma_{\mu}q_r)(\bar{q}_s\gamma^{\mu} q_t)$ &  $c_{uu}$  & $(\bar{u}_p \gamma_{\mu}u_r)(\bar{u}_s\gamma^{\mu} u_t)$ &  $c_{lu}$  & $(\bar{l}_p \gamma_{\mu}l_r)(\bar{u}_s\gamma^{\mu} u_t)$\\
      $c_{qq}^{(3)}$  & $(\bar{q}_p \gamma_{\mu} \tau^I q_r)(\bar{q}_s\gamma^{\mu} \tau^I q_t)$ &  $c_{dd}$  & $(\bar{d}_p \gamma_{\mu}d_r)(\bar{d}_s\gamma^{\mu} d_t)$ &  $c_{ld}$  & $(\bar{l}_p \gamma_{\mu}l_r)(\bar{d}_s\gamma^{\mu} d_t)$\\
        $c_{lq}^{(1)}$  & $(\bar{l}_p \gamma_{\mu}l_r)(\bar{q}_s\gamma^{\mu} q_t)$ &  $c_{eu}$  & $(\bar{e}_p \gamma_{\mu}e_r)(\bar{u}_s\gamma^{\mu} u_t)$ &  $c_{qe}$  & $(\bar{q}_p \gamma_{\mu}q_r)(\bar{e}_s\gamma^{\mu} e_t)$\\
        $c_{lq}^{(3)}$  & $(\bar{l}_p \gamma_{\mu} \tau^I l_r)(\bar{q}_s\gamma^{\mu} \tau^I q_t)$ &  $c_{ed}$  & $(\bar{e}_p \gamma_{\mu}e_r)(\bar{d}_s\gamma^{\mu} d_t)$ &  $c_{qu}^{(1)}$  & $(\bar{q}_p \gamma_{\mu}q_r)(\bar{u}_s\gamma^{\mu} u_t)$\\
         &  &  $c_{ud}^{(1)}$  & $(\bar{u}_p \gamma_{\mu}u_r)(\bar{d}_s\gamma^{\mu} d_t)$ &  $c_{qu}^{(8)}$  & $(\bar{q}_p \gamma_{\mu} T^A q_r)(\bar{u}_s\gamma^{\mu} T^A u_t)$\\
         &  &  $c_{ud}^{(8)}$  & $(\bar{u}_p \gamma_{\mu} T^A u_r)(\bar{d}_s\gamma^{\mu} T^A d_t)$ &  $c_{qd}^{(1)}$  & $(\bar{q}_p \gamma_{\mu}q_r)(\bar{d}_s\gamma^{\mu} d_t)$\\
         &&&&  $c_{qd}^{(8)}$  & $(\bar{q}_p \gamma_{\mu} T^A q_r)(\bar{d}_s\gamma^{\mu} T^A d_t)$
    \end{tabular}
    \caption{The 4-fermion dimension-6 SMEFT operators and corresponding Wilson coefficients in the Warsaw  basis \cite{Grzadkowski:2010es} that do not  explicitly break flavour SU(3).}
    \label{tab:dim6wilsonfermi}
\end{table}

In LHC measurements, cross-sections are measured, which scale with the amplitude of the Lagrangian squared. If we consider Eq.~\eqref{eq:smeftlagrangian}, taking only the dimension-6 amplitude into account, we can infer that the cross-sections measured at the LHC will scale with
\begin{equation}
     xs \approx \mathcal{L}_{\text{EFT}}^2 = \mathcal{L}_{\rm SM}^2 + \mathcal{L}_{\rm SM} \times \frac{1}{\Lambda_{\text{NP}}^2} \sum_i{{c_i}\mathcal{O}_i^{(6)}} + \frac{1}{\Lambda_{\text{NP}}^4} \left(\sum_i{{c_i}\mathcal{O}_i^{(6)}}\right)^2
     \label{eq:dim6scalinglambda}
\end{equation}
where
the dimension-6 squared term is of the same order in $\Lambda_{\text{NP}}$ as the dimension-8 linear terms, i.e., $\frac{1}{\Lambda_{\text{NP}}^4}$. Often, measurements at the LHC show limits on Wilson coefficients both in the linear assumption, meaning ignoring the last term in Eq.~\eqref{eq:dim6scalinglambda} that scales with $\frac{1}{\Lambda_{\text{NP}}^4}$, as well as under the quadratic assumption, meaning this term is not ignored, to give an indication for the size of the contribution of dimension-8. In the case where $\Lambda_{\text{NP}} = M_{\rm Planck} \approx 10^{19}\, \rm GeV$, the suppression of the dimension-8 term is large enough that dimension-8 can be safely ignored for the time being.  An exception to this expectation is if the Wilson coefficients satisfy $c_i\gg \mathcal{O}(1)$, which will be relevant for some of our scenarios below.

\begin{figure}
    \centering
    \includegraphics[width=0.49 \textwidth]{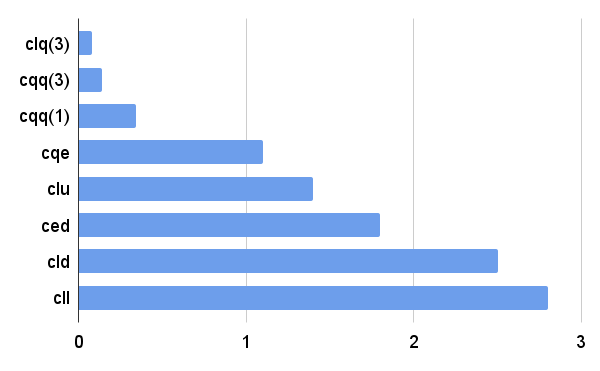}
    \includegraphics[width=0.49 \textwidth]{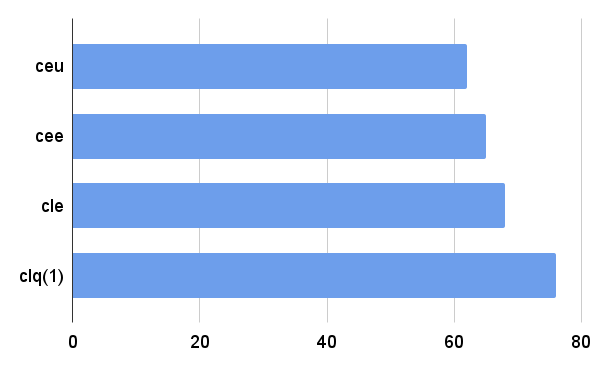}
    \caption{Approximate current experimental limits on four fermion SMEFT parameters showing those parameters constrained to below five (left) and below 100 (right). The values compatible with experiment lie within the blue ranges. For these limits a scale of new physics $\Lambda_{\text{NP}}$ is assumed of 1 TeV. }
    \label{fig:explims}
\end{figure}

Most of the coefficients primarily cause a shift in the Fermi constant, resulting in an overall normalization factor across different analyses. These normalization changes can be measured in experiments such as the ATLAS and CMS experiments at the LHC \cite{ATLASref,CMSref}. Currently no significant deviations from Standard Model (SM) predictions have been found and experimental limits can be set. Approximate current experimental limits on four-fermion interactions under the linear assumption are shown in Fig.~\ref{fig:explims}, \cite{ATLAS:2021kog,Collaboration:2741341,Bellan:2021dcy,CMS:2020pnn}. The Wilson coefficients not shown in this figure are either not measured or limits could not be set to below 100.  The limits can be understood as a limit on the absolute value of the Wilson coefficient. It should be noted that most limits are approximately the same in the positive and negative directions, but they are not by construction symmetric.

\section{Review: Asymptotically safe gravity-matter models and fixed-point structure}
\label{sec:review}
 Asymptotically safe quantum gravity is best understood in the language of the Renormalization Group (RG). Thus, we first introduce some methodology and terminology on the functional RG, before we review the current status of this candidate quantum gravity theory.
\subsection{Flow equation and fixed point}
Physics at a given energy scale can be completely described by the momentum modes at and below this energy scale \cite{Weinberg:1978kz}.  In such an effective theory, the large-momentum modes affect the low-energy observables by altering the parameters of the effective theory. These effects are codified in the flow equation  \cite{Wetterich:1992yh, Morris:1993qb},  adapted to gravity in \cite{Reuter:1996cp},

\begin{equation}\label{eq:masterEquation}
     \partial_t \, \Gamma_k = \frac{1}{2} \operatorname{STr} \left ( \frac{ \partial_t R_k}{\Gamma^{(2)}_k + R_k} \right ) \, , 
\end{equation}
where $\Gamma_k$ is the (Euclidean) scale-dependent effective action where all modes with momenta, $p \equiv (p \cdot p)^{1/2}$ , above $k$ are integrated out, and $\Gamma_k^{(2)}$  denotes the second functional derivative with respect to the field content. $R_k(p^2)$  is a momentum dependent IR regulator that suppresses modes below the scale $k$  by endowing them with a $k$- and momentum-dependent mass term. It also acts as an UV cutoff through its derivative, $\partial_t R_k (p^2)$, with respect to RG time ($t \equiv \log k$). STr denotes the super-trace over all the momentum modes and internal and spacetime indices, including a negative sign for fermionic fields.  This equation captures the flow of the effective action with respect to the scale $k$ by integrating over the momentum modes around a momentum shell $k$. 

Quantum fluctuations generate all possible interactions which respect the symmetries in the theory, thus the effective action can be written as 
\begin{equation}
    \Gamma_k = \sum_j  \, \frac{g_j}{k^{d_{g_j}}} \, \int \mathcal{O}_j \, ,\label{eq:Gammak}
\end{equation}
where $\mathcal{O}_j$ are the  quasi-local operators that preserve the symmetries of the theory and $g_j$ are the  dimensionless couplings strengths corresponding to each of these interactions  and $d_{g_j}$ is the mass-dimension of a coupling. In the limit $k\rightarrow 0$, $\Gamma_k$ reduces to the standard effective action, $\Gamma_{k \rightarrow 0} = \Gamma$, such that, for the SM degrees of freedom, Eq.~\eqref{eq:Gammak} reduces to Eq.~\eqref{eq:smeftlagrangian} and $\frac{g_i}{k^{d_{g_i}}} \rightarrow \frac{c_i}{\Lambda_{\text{NP}}^{d_{g_i}}}$, where $\Lambda_{\text{NP}}$ in the simplest case corresponds to the mass scale of the modes that have been integrated out.

Although it appears that there are innumerable parameters corresponding to each of the operators in the scale-dependent effective action, we often find that only a finite number of them are actually free parameters, which are fixed by a finite number of experiments, and the rest are either determined in terms of these free parameters or their effects can be ignored at a given order of precision.

To keep the computations tractable, it is necessary to truncate the effective action, and ensure the chosen truncation is a good approximation to capture the relevant physics. Within a truncation it can then be determined how many of the couplings correspond to free parameters. By extending the truncation systematically and looking for apparent convergence of fixed-point values, the robustness of results can be assessed and systematic uncertainties estimated.

 From the left hand side of Eq.~\ref{eq:masterEquation}, the beta functions can be extracted, 
\begin{equation}
    \partial_t \Gamma_k \equiv 
    \sum_j \partial_t \frac{g_j}{k^{d_{g_j}}} { \int}{\mathcal{O}_j}\equiv  \sum_j  \left(\frac{\beta_{g_j}}{ k^{d_{g_j}}} - d_{g_j}\frac{g_j}{k^{d_{g_j}}}\right){ \int}{\mathcal{O}_j}\, ,
\end{equation}
which captures the flow of individual couplings. In a UV complete theory, the couplings do not change much above a certain scale, and instead approach a finite fixed point. This requires that all the beta functions vanish simultaneously, $\{ \beta_{g_j} \} = 0$. 

For a given fixed point, we can determine which of the couplings are the free parameters and which are  predictions, by computing the stability matrix
\begin{equation}
    M_{ij} \equiv \frac{\partial \beta_{g_j}}{\partial g_i} |_{g^\ast}  \, ,
\end{equation}
and critical exponents, $\theta_I$, which are the eigenvalues of $M_{ij}$, multiplied by an additional negative sign. The couplings associated with the positive critical exponents are the free parameters, and these couplings are referred to as relevant couplings. As the flow of such couplings quickly departs from their fixed-point value as the scale decreases, this corresponds to an IR repulsive flow. The couplings associated with negative critical exponents correspond to predictions, and are referred to as irrelevant couplings. In contrast to the relevant couplings, the flow of the irrelevant couplings {(towards lower energy scales)} is pulled closer to the fixed-point value even if it is perturbed. The flow of all the couplings from an UV fixed point down to an IR scale is completely determined by the choice of relevant coupling values close to the fixed point, which is based on experimentally measured IR values.

\subsection{Asymptotically safe gravity and matter}

The Einstein-Hilbert action is perturbatively non-renormalizable, as it generates divergences for
infinitely many higher-dimension operators.   
Each divergence requires a counterterm and each counterterm comes with a free parameter; thus the theory has the same status as the SMEFT in that one can make predictions at sufficiently low energies (see \cite{Donoghue:2022eay} for a review), but ultimately expects new physics to appear. 
Traditionally, the new physics has been assumed to lie outside the realm of standard local quantum field theories, and come either in the form of a non-local theory like string theory or a fundamentally discrete theory like Loop Quantum Gravity or causal set theory.
However, there is increasingly compelling evidence, based on non-perturbative methods such as the FRG, that 
gravity is asymptotically safe \cite{Lauscher:2001ya,Reuter:2001ag,Litim:2003vp,Codello:2006in,Codello:2007bd,Machado:2007ea,Codello:2008vh,Benedetti:2009rx,Benedetti:2010nr,Eichhorn:2010tb,Manrique:2011jc,Falls:2013bv,Codello:2013fpa,Christiansen:2014raa,Becker:2014qya,Christiansen:2015rva,Ohta:2015efa,Gies:2016con,Biemans:2016rvp,Denz:2016qks,Falls:2017lst,Gonzalez-Martin:2017gza,Knorr:2017mhu,Christiansen:2017bsy,Bosma:2019aiu,Knorr:2019atm,Falls:2020qhj,Kluth:2020bdv,Knorr:2020ckv,Bonanno:2021squ,Fehre:2021eob,Knorr:2021slg,Knorr:2021niv,Baldazzi:2021orb,Mitchell:2021qjr,Sen:2021ffc,Kluth:2022vnq,Knorr:2023usb,Saueressig:2023tfy,Baldazzi:2023pep,Becker:2024tuw}, where all the couplings associated with the higher-dimension operators have a 
UV fixed point and gravity can be quantized as a standard QFT. There is mounting evidence that this fixed point, the Reuter fixed point, comes 
with only three relevant parameters \cite{Codello:2007bd,Benedetti:2009rx,Falls:2013bv,Denz:2016qks,Gies:2016con,Falls:2017lst,Falls:2020qhj,Becker:2024tuw}, and the rest are irrelevant and are determined in terms of these three  parameters. In the context of the EFT for gravity this implies that all but three couplings of the theory are calculable in terms of the three free parameters, which are the Newton coupling, cosmological constant and a curvature-squared coupling \cite{Gubitosi:2018gsl}.

Since matter couples to gravity, it alters the flow of the gravitational couplings; conversely gravitational fluctuations also affect the flow of matter  couplings. Therefore, a fixed point found in the pure gravity system might not persist when particular matter models are coupled to it.
There is mounting evidence that the fixed point in the gravitational couplings persists if we include matter content of different models such as the Standard Model and many of its  extensions \cite{Narain:2009fy,Eichhorn:2011pc,Dona:2012am,Dona:2013qba,Dona:2014pla,Meibohm:2015twa,Oda:2015sma,Percacci:2015wwa, Labus:2015ska, Meibohm:2016mkp,Dona:2015tnf,Eichhorn:2016vvy,Biemans:2017zca,Christiansen:2017cxa,Alkofer:2018fxj,Eichhorn:2018ydy,Eichhorn:2018nda,Bonanno:2018gck,Wetterich:2019zdo, Eichhorn:2019dhg, DeBrito:2019gdd, Daas:2020dyo,  Ali:2020znq, Laporte:2021kyp, Sen:2021ffc,Daas:2021abx,Schiffer:2021gwl}. Moreover, non-minimal interactions between gravity and matter are often generated \cite{Eichhorn:2017sok, Eichhorn:2018nda,Laporte:2021kyp, Knorr:2024yiu}.
Furthermore, there are  strong indications that interactions in the matter sector can feature a fixed point under the impact of gravity fluctuations \cite{Shaposhnikov:2009pv,Harst:2011zx,Eichhorn:2017ylw,Eichhorn:2017als,Eichhorn:2017muy, Eichhorn:2017lry,Eichhorn:2018whv,Pawlowski:2018ixd,DeBrito:2019gdd,Eichhorn:2019dhg,Eichhorn:2020sbo}.  At such a fixed point, matter couplings can correspond to irrelevant directions, strengthening the predictive power of asymptotically safe gravity-matter models. This can have important phenomenological consequences. First, some couplings, most notably the Yukawa couplings  and the Abelian gauge coupling of the SM, feature upper bounds \cite{Eichhorn:2017ylw, Eichhorn:2017lry,Eichhorn:2018whv}. Second, some matter models, e.g., specific proposals for dark matter, are not UV complete \cite{Eichhorn:2017als,deBrito:2023ydd} or strongly constrained in their parameter values \cite{Reichert:2019car,Hamada:2020vnf,Eichhorn:2020kca,Eichhorn:2020sbo,Kowalska:2020zve}.  These results open the door for observational tests of asymptotically safe gravity.

In the context of the present work, it is particularly important that the effect of gravity on matter is not limited to modifying the beta functions of perturbatively renormalizable matter couplings. Instead, gravity also generates higher-order matter interactions already at the UV fixed point \cite{Eichhorn:2012va, Eichhorn:2016esv, Eichhorn:2017eht, Eichhorn:2017sok, Eichhorn:2018nda, Eichhorn:2021qet,deBrito:2021pyi,deBrito:2023myf}. Intriguingly, these are exactly the type of interactions included in the SMEFT, because they are perturbatively non-renormalizable, quasi-local interactions. For instance, these include four-fermion couplings \cite{Eichhorn:2011pc, Meibohm:2016mkp,Eichhorn:2017eht, deBrito:2020dta,deBrito:2023kow}, which will be our focus below.\footnote{Other candidate quantum-gravity theories, such as quadratic gravity, also make predictions for four-fermion couplings \cite{deBrito:2023pli}.} The main features of asymptotic safety when it comes to these couplings are that there is generically a fixed point with non-zero values of these couplings, at which these couplings are irrelevant and therefore their low-energy values can be calculated. In other words, (some) 
higher-order couplings
in the SMEFT are expected to be nonzero in asymptotic safety. 
We will investigate the consequences of this for experimental constraints on the SMEFT couplings below.  Along a related line of research, a comparison to positivity bounds on higher-order photon-interactions has recently been done for the first time \cite{Knorr:2024yiu,Eichhorn:2024wba}. More generally, a first step to mapping out the asymptotically safe ``landscape"\footnote{A comparison to the string-theoretic landscape is discussed in \cite{Eichhorn:2024rkc}.} was done in \cite{Basile:2021krr,Knorr:2024yiu} and a calculational strategy has been developed in \cite{Saueressig:2024ojx}.

\section{Toy model: Four-fermions interactions}\label{sec:fourfermionresults}
\subsection{Action and field content}
Following \cite{Eichhorn:2011pc}, we shall consider a model defined by the following Euclidean action, as a proxy for the full SMEFT  with gravity:
\begin{align}
    S &= S_\text{EH} + S_\text{kin,F} + S_{4\text{F}} \\
    S_\text{EH} &= \frac{1}{16\pi \bar{G}_\mathrm{N}} \int_x \sqrt{g} (2 \bar{\Lambda} - R) \\
    S_\text{kin,F} &= \int_x \sqrt{g} \bar\psi \slashed{\nabla} \psi \\
    S_{4\text{F}} &= \int_x \sqrt{g} \left[ \frac{\bar\lambda_+}{2} \left(\mathcal{V} + \mathcal{A}\right) + \frac{\bar\lambda_-}{2} \left(\mathcal{V} - \mathcal{A}\right) \right]
\end{align}
The first piece is the Einstein--Hilbert action for the metric tensor $g_{\mu\nu}$; $\bar{G}_\text{N}$ and $\bar\Lambda$ denote the (dimensionful) Newton coupling and cosmological constant respectively, and $R$ denotes the Ricci scalar associated with $g_{\mu\nu}$. Fluctuations of the spacetime metric are parametrized by setting
\begin{align}
    g_{\mu\nu} = \delta_{\mu\nu} + \sqrt{\bar{G}_\text{N}} h_{\mu\nu}.
\end{align}
The factor $\sqrt{\bar{G}_\text{N}}$ ensures that $h_{\mu\nu}$ has mass dimension 1, as should any bosonic field with an inverse propagator quadratic in momentum. Note that momentum is a good quantum number, since we are expanding about a flat background $\bar{g}_{\mu\nu} = \delta_{\mu\nu}$.

The matter sector consists of $N_\text{f}$ Dirac fermions. These are coupled minimally to metric fluctuations via the metric determinant in the volume measure and the covariant derivative $\nabla_\mu$, which contains the spin connection. The piece $S_{4\text{F}}$ contains the two four-fermion operators
\begin{align}\label{Eq:fourFermionTermsDiracBasis}
    \mathcal{V} \pm \mathcal{A} = \left(\bar\psi \gamma_\mu \psi\right)^2 \mp \left(\bar\psi \gamma_\mu \gamma_5 \psi\right)^2.
\end{align}
Assuming an $\operatorname{SU}(N_\text{f})_\text{L} \times \operatorname{SU}(N_\text{f})_\text{R}$ chiral symmetry, $S_{4\text{F}}$ constitutes a full Fierz-complete basis.  This symmetry is significantly larger than that of the SM; essentially, we are imposing universality of 4-Fermi interactions with respect to weak and colour isospin, as well as generation.

 In terms of the Warsaw basis, this amounts to setting $c_{\cdot \cdot}^{(3)} = 0$, $c_{\cdot \cdot}^{(8)} = 0$ and for the remaining $c_i$'s (including the $c_{\cdot \cdot}^{(1)}$'s)
\begin{align}
    \frac{c_{i}}{M_\text{Pl}^2} &= \bar\lambda_- \qquad (i \in \{ (\bar{L}L)(\bar{L}L), (\bar{R}R)(\bar{R}R) \}) \\
    \frac{c_{i}}{M_\text{Pl}^2} &= 2\bar\lambda_+ \qquad (i \in (\bar{L}L)(\bar{R}R))
\end{align}
in the limit $k \to 0$.\footnote{Because the $c_i$ by definition do not depend on the scale, there is no possibility for a scale-dependent identification. Instead, $\lambda_{\pm}$ can only be equated to $c_i$'s in the limit $k \rightarrow 0$.} Assuming no further matter degrees of freedom, $S_\text{kin,F} + S_{4\text{F}}$ constitutes a full basis up to dimension 6.

\subsection{Beta functions}
The  gravitational contributions to the beta functions for the dimensionless matter couplings $\lambda_\pm = k^2 \bar{\lambda}_\pm$ were derived in Ref.~\cite{Eichhorn:2011pc}, see also \cite{deBrito:2020dta,deBrito:2023kow} for more general settings; the pure-matter contribution was first derived in \cite{Gies:2003dp}. The beta functions read
\begin{equation}
	\begin{aligned}
		\beta_{\lambda_+} &= ( 2 \, +  \eta_{
   4\rm F
  }) \,\lambda_{+} 
		+ \frac{3\lambda_{+}^2}{8\pi^2} + \frac{(1+ \Nf) \, \lambda_{+} \lambda_{-}}{4\pi^2} + \frac{5 \,G^2}{8 (1-2 \Lambda )^3} ,
	\end{aligned}
\end{equation}
		
\begin{equation}
	\begin{aligned}
		\beta_{\lambda_-} &= ( 2 \, +  \eta_{
  4\rm F
  }) \,\lambda_{-}  \!-\! \frac{(1- \Nf) \, \lambda_{-}^2}{8 \pi ^2} \!+\! \frac{\Nf \, \lambda_{+}^2}{8 \pi ^2} - \frac{5 \, G^2}{8 (1-2 \Lambda )^3}.
	\end{aligned}
\end{equation}
Here, $G = k^2 \bar{G}_\text{N}$ denotes the dimensionless Newton coupling and $\Lambda = 
k^{-2}
\bar{\Lambda}$ the dimensionless cosmological constant. Because gravity couples to the spacetime indices, not the internal indices of matter fields, it is `blind' to internal symmetries. Therefore, the gravitational contribution to the scaling dimension at a free fixed point of all 4-fermion channels is the same, i.e., there is a Fierz-universality of the gravitational contribution, see \cite{Eichhorn:2023jyr} for more details. This contribution is denoted as $\eta_{4\text{F}}$, and reads
\begin{equation}\label{eq:eta4ferm}
	\eta_{4\text{F}} = 2 \eta_\psi +  \frac{5\,G}{2 \pi  (1-2 \Lambda )^2} - \frac{G}{20 \pi  (1 - 4 \Lambda/3)} - \frac{31\,G}{60 \pi  (1 - 4 \Lambda/3)^2} \,.
\end{equation}
Here, $\eta_\psi$ is the one-loop fermion anomalous dimension, and receives contributions only from gravity, as in \cite{deBrito:2020dta}:
\begin{equation}
	\eta_\psi = \frac{3\, G}{20 \pi  (1-4 \Lambda/3)} - \frac{25 \,G}{16 \pi  (1-2 \Lambda )^2} + \frac{29\,G}{80 \pi (1-4 \Lambda/3)^2} \,.
\end{equation}

\subsection{Fixed-point structure}
\begin{figure}
    \centering
    \includegraphics[width=0.48\textwidth]{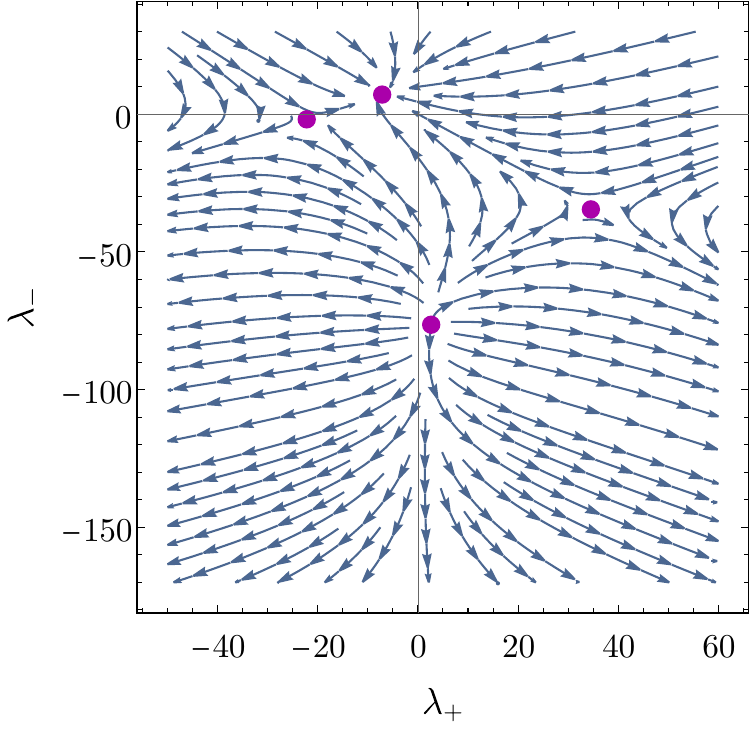} \hfill \includegraphics[width=0.48\textwidth]{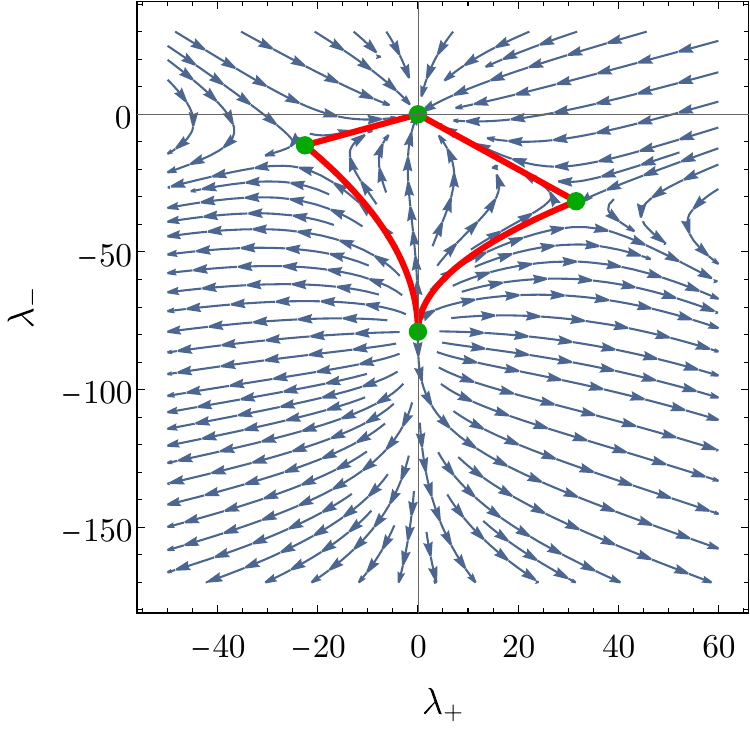}
    \caption{Fixed-point structure and RG phase portrait with (left) and without (right) gravity contributions, the former for fiducial values of the gravitational couplings $(G_*,\Lambda_*) = (5,0)$. The red-demarcated region denotes the part of theory space attracted simultaneously to the Gaussian fixed point in the IR and the UV attractive fixed point in the UV.}
    \label{fig:FPs}
\end{figure}

This set of beta functions leads to the RG phase portrait shown in Fig.~\ref{fig:FPs}(a) for fiducial fixed-point values of the gravitational couplings. Below the Planck scale, the dimensionless Newton coupling decays quickly to zero. Consequently, the RG evolution can be captured to a good approximation by setting $G(k < M_\text{Pl}) = 0$ in the beta functions above. It is then instructive to also keep in mind the RG phase portrait for $G = 0$, which is shown in Fig.~\ref{fig:FPs}(b). It was first derived in \cite{Gies:2003dp} and consists of four fixed points. First, there is the Gaussian fixed point. It  has two irrelevant directions, i.e., both four-fermion couplings are attracted to it under the RG flow. Second, there are three interacting fixed-point candidates. Two of them have one irrelevant and one relevant direction, and are thus attractive in one direction and repulsive in another direction.  We call them metastable. As is visible from Fig.~\ref{fig:FPs}(b), none of these directions is aligned with any of the two couplings. The third fixed-point candidate has two relevant directions, i.e., it cannot be the IR endpoint of an RG flow, unless the theory is scale-invariant and thus described by the fixed-point candidate at all scales.  This fixed point is called unstable. In the absence of gravity, these fixed points are, depending on context, either called Gross--Neveu-like or Nambu--Jona-Lasinio-like fixed points.

In this characterization of the phase diagram, we have been careful to distinguish the Gaussian fixed point from the interacting fixed-point candidates. The latter, due to the appearance of new relevant directions, would require an in-depth study with a larger truncation of the full dynamics to robustly establish their existence.\footnote{In $d<4$, their existence is much more robustly established, see Refs.~\cite{Zerf:2017zqi,Gracey:2016mio}, \cite{Gracey:2021ili}, \cite{Knorr:2016sfs} and \cite{Iliesiu:2017nrv} for state-of-the-art $\epsilon$ expansion, large-$N$ expansion, FRG and conformal bootstrap respectively.} For our purposes, however, it is not important whether these fixed points exist beyond the present toy-model setting, because we only use our toy model to discuss three different, in principle possible, scenarios for the implications of asymptotically safe gravity for the SMEFT. Any statement about the actual SMEFT has to be made in a setting that accounts for all degrees of freedom in the SMEFT, and thus the robust establishment of fixed points is a task deferred to a much more involved study than the present one.

Under the impact of quantum-gravity fluctuations, the phase portrait gets deformed, but is not changed qualitatively. As discussed in Sec.~\ref{sec:review}, the IR attractive fixed point (i.e., the shifted Gaussian fixed point\footnote{We adopt the terminology of \cite{Eichhorn:2011pc}, in which a fixed point which is interacting, but becomes the Gaussian fixed point when the gravitational coupling $G$ is set to zero, is called the shifted Gaussian fixed point.}) in the transplanckian regime features non-vanishing 4-Fermi couplings, generated purely by fluctuations of the spacetime metric. However, its scaling spectrum is close (at least for $G$ small enough) to that of the Gaussian fixed point (GFP), which exists for $G=0$ and is IR attractive in that case. In addition, the phase portrait still features two meta-stable and one unstable (towards the IR) fixed-point candidates.

\section{Scenarios and illustrative trajectories}\label{sec:scenarios}
Generically, one expects that any interaction with a coupling of dimension $-d_g$ that is generated by quantum gravity is suppressed by $\left(\frac{E}{M_{\rm Pl}}\right)^{d_g}$, which would imply a quadratic suppression with energy for four-fermion couplings. Below, we show that this expectation holds true for the most conservative (and most perturbative) choice of fixed point, but may not hold for fixed points that are non-perturbative in nature.

\subsection{Shifted Gaussian fixed point}
The sGFP \emph{per se} defines a highly predictive universality class. Since it is IR attractive, the only UV safe trajectory is fixed by $\lambda_\pm(k > M_\text{Pl}) = \lambda_{\pm,*}|_\text{sGFP}$. Given fixed-point values $\lambda_{\pm,*}|_\text{sGFP}$, one could in principle derive quantitative predictions for the corresponding Wilson coefficients at LHC scales, which could be compared to measurements. However, the fixed-point values in practice are not known with sufficient precision. Qualitatively, on the other hand, the sGFP's universality class is no different from the SMEFT without gravity. In particular, since $|\lambda_{\pm,*}|_\text{sGFP}| \sim 1$ (more precisely, it is $O(G_*)$), one has
\begin{align}
    \left|\lambda_\pm(k = M_\text{LHC})\right| \sim \left(\frac{M_\text{LHC}}{M_\text{Pl}}\right)^2.
\end{align}

\subsection{Fixed point with relevant directions at transplanckian scales}
If a UV attractive fixed point exists in the transplanckian regime, then the corresponding universality class is one where the values of the couplings at the Planck scale, $\lambda_\pm(k=M_\text{Pl})$, are free parameters of the theory. If the flow without gravity is quasi-classical, one would still approximately have $\lambda_\pm(k=M_\text{LHC}) \approx \lambda_\pm(k=M_\text{Pl})\left(M_\text{LHC}/M_\text{Pl}\right)^2$. However, one could set, e.g.,
\begin{align}
    |\lambda_\pm(k=M_\text{Pl})_{\text{Scen 2A}}| = \left(\frac{M_\text{Pl}}{M_\text{LHC}}\right)^\delta
\end{align}
for some $\delta > 0$ or
\begin{align}
    |\lambda_\pm(k=M_\text{Pl})_{\text{Scen 2B}}| = \left(\frac{M_\text{Pl}}{M_\text{NP}'}\right)^2,
\end{align}
leading to violation of naturalness expectations of the form
\begin{align}
    |\lambda_\pm(k=M_\text{Pl})_{\text{Scen 2A}}| &\sim \left(\frac{M_\text{LHC}}{M_\text{Pl}}\right)^{2-\delta},\\
    |\lambda_\pm(k=M_\text{Pl})_{\text{Scen 2B}}| &\sim \left(\frac{M_\text{LHC}}{M_\text{NP}'}\right)^2.
\end{align}
The former mimics a violation of classical scaling without changing the subplanckian dynamics significantly, whilst the latter mimics the existence of a non-gravitational `New Physics' scale. This unnatural choice of Planck-scale data does \emph{not} incur a Landau pole in the UV---in other words, the RG trajectories remain UV-complete---owing to the existence of a UV attractive fixed point. There are no general arguments why a scenario of this kind should not exist. However, we find that in our system of beta functions, trajectories of this kind are not attracted towards the infrared by the Gaussian fixed point below the Planck scale. Rather, they flow off to infinity within finite RG `time', signalling the onset of chiral symmetry breaking due to strong coupling. If we wish QCD and/or electroweak symmetry breaking to be the sole source(s) of chiral symmetry breaking, this is an unattractive scenario on phenomenological grounds. If we initiate the RG flow within the IR basin of attraction of the Gaussian fixed point and simultaneously close to a UV attractive fixed point, we in fact observe the opposite phenomenon: (comparatively) natural Planck-scale data already mimic a distinct `New Physics' scale (Scenario 2B above), because they pass close to a non-trivial subplanckian fixed point. This is the subject of Scenario 3 below.

\subsection{Interacting fixed points above and below the Planck scale}
\begin{figure}
    \centering
    \begin{minipage}{.45\textwidth}(a)\\[-1em]
        \includegraphics[width=\textwidth]{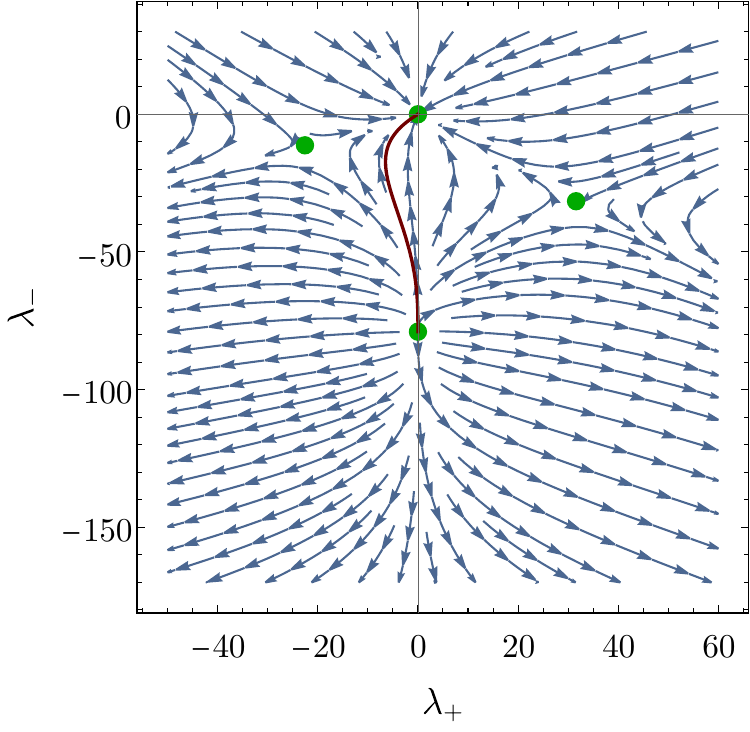}\\[1em]
        (b)\\[-1em]
        \includegraphics[width=\textwidth]{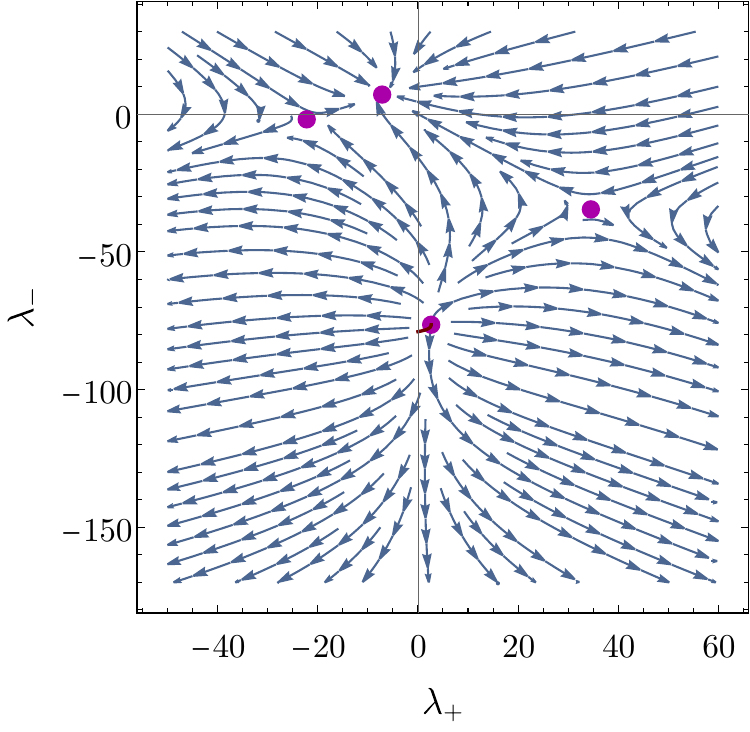}
    \end{minipage}
    \hfill
    \begin{minipage}{.50\textwidth}
        (c)\\[-.5em]
        \includegraphics[width=\textwidth]{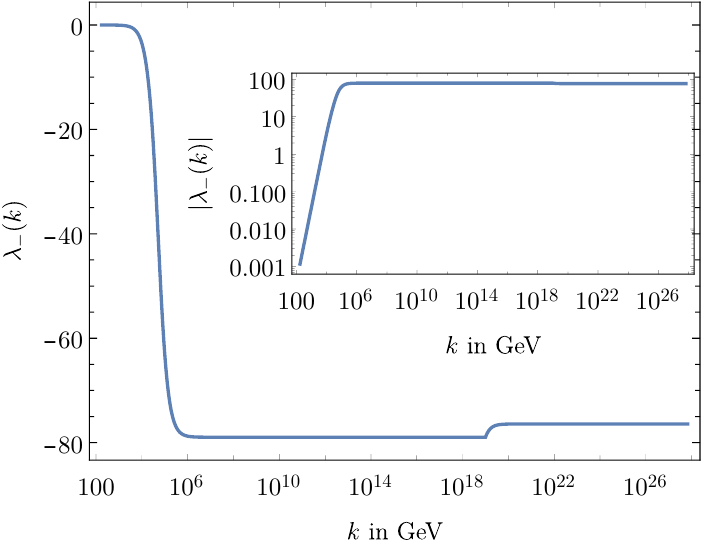}
        \\[1em]
        (d)\\[-.5em]
        \includegraphics[width=\textwidth]{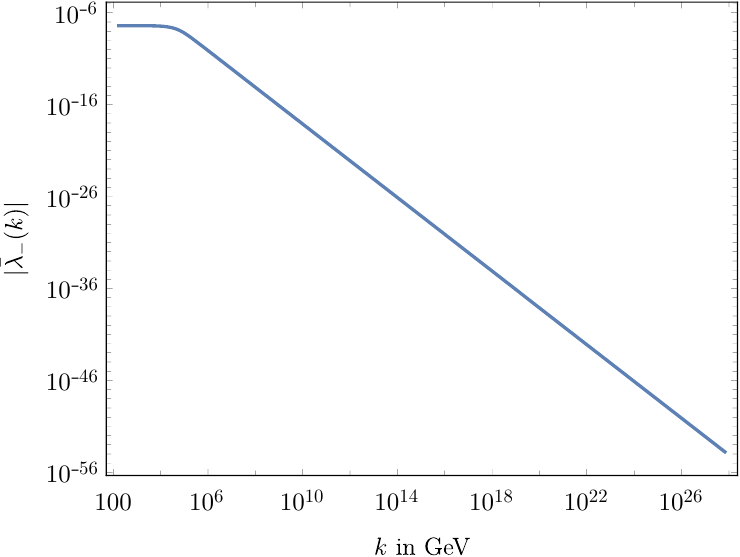}
    \end{minipage}
    \caption{Sample trajectory that allows the realisation of 4-Fermi couplings that deviate from naturalness expectations, with $|\lambda_\pm(k=M_\text{LHC})| \sim 10^{-3} \gg (M_\text{LHC}/M_\text{Pl})^2$. Left panels: Flow in theory space, with the sub- and transplanckian regimes displayed in (a) and (b) respectively. Right panels: Flow of dimensionless (c) and dimensionful (d) running couplings as a function of RG scale $k$, with the double logarithmic inset in (c) showing the onset of classical power-law running well below the Planck scale.}
    \label{fig:sampletrajectory}
\end{figure}
In this scenario, the flow below the Planck scale spends a long `RG time duration' in the vicinity of the subplanckian UV attractive fixed point. Instead of the Planck mass, there is a different effective New Physics scale $M_\text{non-pert}$ arising from the non-perturbative mechanism needed to generate the subplanckian fixed point. The New Physics scale is not tied to new degrees of freedom, as one would conventionally expect, but rather to an onset of nonperturbative interactions between the already existing degrees of freedom. As in Scenario 2 above, the scenario is nevertheless UV complete, due to a nonperturbative matter-gravity fixed point into which the reverse RG flow is attracted.
The IR value of the 4-Fermi coupling roughly follows naturalness expectations, but only if the new-physics scale is re-interpreted:
\begin{align}
    |\lambda_\pm(k = M_\text{LHC})| \sim \left(\frac{M_\text{LHC}}{M_\text{NP}}\right)^2 = \left(\frac{M_\text{LHC}}{M_\text{non-pert}}\right)^2.
\end{align}
A sample trajectory is shown in Fig.~\ref{fig:sampletrajectory}. These trajectories are in fact \emph{generic}, as long as one imposes boundary values $\lambda_\pm(k=M_\text{LHC})$ that lie within the basin of attraction of both the Gaussian fixed point with respect to flows towards the IR and the UV attractive fixed point for flows towards the UV. At the present level of approximation, this can be computed analytically and found to be the region enclosed by the four curves
\begin{align}
    \mathcal{S}_{N,Q} &= \left\{(1-t) \vec{\lambda}_*|_N + t \vec{\lambda}_*|_{Q} \right\}_{t \in [0,1]}, \\
    \mathcal{S}_{B,Q} &= \left\{((1-t) \lambda_{+,*}|_B + t^2 \lambda_{+,*}|_Q, (1-t) \lambda_{-,*}|_B + t \lambda_{-,*}|_Q ) \right\}_{t \in [0,1]}
\end{align}
Here $Q$ denotes any of the two mixed-stability fixed points, $N$ the Gaussian fixed point and $B$ the UV attractive fixed point (all evaluated at $G=0$). This yields the red-demarcated region shown in the right panel of Fig.~\ref{fig:FPs}(b).

\section{Conclusions and outlook}\label{sec:conclusions}
It is a crucial task to connect quantum gravity to experiments. This is often viewed as extremely challenging, due to the gap in scales between the Planck scale and scales accessible, e.g., at the LHC. Yet, under the assumption of no new physics between the LHC scales and the Planck scale\footnote{By ``no new physics" we more precisely mean no new physics which is strongly coupled enough to change the RG flow by $\mathcal{O}(1)$ effects.}, the RG flow provides a direct mapping of Planck-scale predictions to LHC scales. In the present paper, we have explored what form this mapping takes for asymptotically safe quantum gravity and the higher-order interactions in the SMEFT. In asymptotically safe gravity, many higher-order interactions in the SMEFT are expected to come with calculable values of their couplings at the Planck scale. These result in calculable values at LHC scales.
We outline three scenarios, which we support by calculations in a toy model for the four-fermion interactions in the SMEFT:

\begin{enumerate}
\item In the most conservative scenario, the Wilson coefficients of four-fermion operators are 
$\sim \left(\frac{M_{\rm LHC}}{M_{\rm Pl}}\right)^2 
$
i.e., they are unmeasurably small.
Extrapolating from our toy model to the SMEFT, this constitutes a testable prediction of this scenario: to the best achievable experimental accuracy of the LHC experiments, no deviations of the SMEFT coefficients from their values with just SM fields is expected.

\item In a less conservative scenario, a non-perturbative UV fixed point is realized at which four-fermion interactions are relevant perturbations. Accordingly, their Planck-scale values can become $\mathcal{O}\!\left((M_{\rm Pl}/M_{\rm LHC})^2\right)$, such that their low-energy value becomes 
$\mathcal{O}(1)$. We do not expect that this scenario is ultimately realized in a viable asymptotically safe theory of gravity and matter, but it may be realized in our toy model and we discuss it to provide a complete picture of the various alternatives.

\item In another less conservative scenario, there is a nonperturbative fixed point at sub-Planckian scales, i.e., just with the SM degrees of freedom. The RG trajectory connecting the UV regime to LHC scales slows down considerably in the fixed point's vicinity, so that the natural suppression of the Wilson coefficients is not realized. Whether or not such a scenario can be realized in the full SM depends on the presence of nonperturbative fixed points which, in turn, require fully non-perturbative studies of the SM at high scales.
\end{enumerate}

Based on our exploration of three scenarios, testable predictions from quantum gravity for the SMEFT are achievable -- even if the most conservative prediction simply amounts to the LHC experiments \emph{not} measuring any deviation of the SMEFT coefficients from their values without new physics. This warrants a more in-depth study that goes beyond the toy model for four-fermion interactions and explores the effect of asymptotically safe quantum gravity on the four-fermion couplings of the SM. Based on their large number, a full study appears out of reach for now, but, e.g., a focus on four-fermion interactions which involve only the quark sector of the third generation (i.e., the heaviest quarks) is achievable by extending the studies in \cite{deBrito:2023kow}.

 When extracting experimental constraints on the four-fermion couplings in the SMEFT, dimension-eight operators and dimension-six operators can mix. 
Dimension-eight operators are suppressed by a factor $1/\Lambda_{\text{NP}}^2$ compared to dimension-six operators and are thus negligible for a new-physics scale close to the Planck scale, under the assumption that all $c_i$ are of the same order of magnitude. 
However, in our work we find two scenarios in which Wilson coefficients are strongly enhanced and in these scenarios, dimension-eight operators might not be negligible.
To compare predictions from asymptotically safe gravity with LHC data, an understanding of dimension-eight operators in asymptotic safety may thus be necessary. Without performing any explicit calculations, we can already state that many dimension-eight operators will have nonzero couplings in the IR. This is because, quantum-gravity fluctuations also generate dimension-eight operators, but they remain irrelevant perturbation of the shifted Gaussian fixed point (if it exists). In principle, the inclusion of a subset of dimension-eight operators in the predictions for scattering amplitudes is possible along the same lines as for dimension-six operators; but a full calculation that accounts for all operators in the SMEFT appears out of reach for now. In practise, accounting for a subset of dimension-eight operators may be feasible.\\

\noindent In summary, we find that under the assumption of a ``desert" between the quantum-gravity scale and the LHC scale, there may be more than one way to ``cross the desert": if our results carry over from our toy model to the SMEFT, distinct scenarios have different implications for the size of Wilson coefficients. All three scenarios have in common that concrete predictions from quantum gravity for the Wilson coefficients can be made and thus tested at the LHC. In two of these scenarios the traditional assumption that quantum gravity induces unmeasurably small Wilson coefficients is not borne out. Thus, quantum-gravity effects may even result in measurable changes of the Wilson coefficients at the LHC.

\acknowledgments
We thank Gustavo P.~de Brito for discussions during the initial stages of this project. We are also grateful to Frank Saueressig for the organization of a Focus Session on Quantum Gravity and the SMEFT at the NWO Physics meeting 2024.
This work is supported by a research grant (29405) from VILLUM FONDEN. S.~R.~further acknowledges support from the Deutsche Forschungsgemeinschaft (DFG) through the Walter Benjamin program (RA3854/1-1, Project id No. 518075237).

\bibliography{references.bib}
\end{document}